# Tunable vortex Majorana zero modes in LiFeAs superconductor

Lingyuan Kong[1†], Lu Cao[1,2†], Shiyu Zhu[1†], Michał Papaj[3†], Guangyang Dai[1,2], Geng Li[1,2], Peng Fan[1,2], Wenyao Liu[1,2], Fazhi Yang[1,2], Xiancheng Wang[1], Shixuan Du[1,2,4], Changqing Jin[1,2,5], Liang Fu[3], Hong-Jun Gao[1,2,4*] and Hong Ding[1,4,5*]

[1]Beijing National Laboratory for Condensed Matter Physics and Institute of Physics, Chinese Academy of Sciences, Beijing 100190, China

[2]School of Physical Sciences, University of Chinese Academy of Sciences, Beijing 100190, China

[3]Department of Physics, Massachusetts Institute of Technology, Cambridge, Massachusetts 02139, USA

[4]CAS Center for Excellence in Topological Quantum Computation, University of Chinese Academy of Sciences, Beijing 100190, China

[5]Songshan Lake Materials Laboratory, Dongguan, Guangdong 523808, China

†These authors contributed equally to this work
*Correspondence to: dingh@iphy.ac.cn, hjgao@iphy.ac.cn

**The recent realization of pristine Majorana zero modes (MZMs) in vortices of iron-based superconductors (FeSCs)[1-5] provides a promising platform for long-sought-after fault-tolerant quantum computation[6-15]. A large topological gap between the MZMs and the lowest excitations enabled detailed characterization of vortex MZMs in those materials[16-23]. Despite those achievements, a practical implementation of topological quantum computation based on MZM braiding[2,24] remains elusive in this new Majorana platform. Among the most pressing issues are the lack of controllable tuning methods for vortex MZMs and inhomogeneity of the FeSC Majorana compounds that destroys MZMs during the braiding process[25]. Thus, the realization of tunable vortex MZMs in a truly homogeneous compound of stoichiometric composition and with a charge neutral cleavage surface is highly desirable. Here we demonstrate experimentally that the stoichiometric superconductor LiFeAs[26] is a good candidate to overcome these two obstacles. Using scanning tunneling microscopy, we discover that the MZMs, which are absent on the natural surface[27], can appear in vortices influenced by native impurities. Our detailed analysis and model calculations clarify the mechanism of emergence of MZMs in this material, paving a way towards MZMs tunable by controllable methods such as electrostatic gating. The tunability of MZMs in this homogeneous material offers an unprecedented platform to manipulate and braid MZMs, the essential ingredients for topological quantum computation.**

Besides the large topological gap, a common advantage of FeSC Majorana platforms, LiFeAs has homogeneous electronic properties (Fig. S3b, S3c) owing to its dopants-free stoichiometric bulk and charge neutral cleavage surface in between the lithium double layers (Fig. 1a). As shown in Fig. 1c, most of the area on the as-cleaved surface of LiFeAs is clean and uniform with an ordered square lattice. Sporadically, some native impurities form spontaneously, likely to reduce the surface energy and are distributed sparsely over the intrinsically homogenous background[28]. Recently, angle-resolved photoemission spectroscopy (ARPES) study reported multiple topological bands in LiFeAs[29]. Two Dirac cones were observed close to the Fermi level, which are the Dirac surface states of topological insulator (TI) phase and the bulk Dirac fermion of topological Dirac semimetal (TDS) phase, respectively (Fig. 1b). Consequently, altering the chemical potential ($\mu$) in LiFeAs is expected to change the topological phase of the material, depending on the nature of the underlying bands incorporated in the vortex quasiparticle excitations. It was predicted that the both topological phases can lead to Majorana quasiparticle in a vortex core. However, the exact type of realized excitation, either the localized single zero modes or mobile helical modes, and the possible interactions between them are highly dependent on the circumstances. The most important factors are the relative positions of the chemical potential and the two topological band crossing points[30,31]. Therefore, the rich topological band structure near the Fermi level opens up a fertile playground for studying Majorana physics in both TI and TDS phases. More importantly however, it provides a rare opportunity to tune vortex MZMs in a homogeneous material by controlling impurities or gating voltage and deformation, as discussed below.

Previous scanning tunneling microscopy/spectroscopy (STM/S) measurements on LiFeAs showed absence of vortex MZM[27], which constituted a major puzzle in the field of vortex-based Majorana platforms[29]. To introduce a tuning parameter necessary in this largely homogeneous system, we took advantage of sparsely distributed impurities as means to enable the emergence of the MZMs[28]. To investigate the morphology of vortex bound states experimentally, we performed low-temperature ($T_{exp}$ = 400 mK), high-resolution (better than 0.28 meV) STM/S measurements on the surface of LiFeAs superconductor. The bare surface after *in-situ* cleavage is formed by lithium atoms. We observe atomically resolved square lattice of lithium atoms in a sizable region free of impurities (Fig. 1c). Two superconducting gaps ($\Delta_1$ = 2.7 meV; $\Delta_2$ = 5.8 meV) are clearly observed under zero field (Fig. 1d), consistent with the previous STM/S results[27,32] (see detailed gap determination in supplementary information). Besides the clean region, some spontaneously formed native impurities are also observed on the cleaved surface (Fig. 1c). With a 2 T magnetic field applied perpendicular to the sample surface, we find that the vortices not only appear as free vortices (*free*Vs) in the clean regions, but also appear as impurity-assisted vortices (*imp*Vs) pinned to the positions of the sparsely distributed impurities (see a schematic illustration in Fig. 1c). Besides an ordinary non-zero-energy peaks observed in all *free*Vs (Fig. 1e), a pronounced zero-bias conductance peak (ZBCP) emerges in some *imp*Vs (Fig. 1f). In addition, the ZBCP is usually accompanied by a pair of energy-symmetric side peaks which are located at about ±0.9 meV, while the spectrum recovers the superconducting gap feature at the *imp*V edges (Figs. S1a-S1b). These phenomena are strikingly similar to the MZM and accompanied integer-quantized vortex bound states observed in the two identified FeSC Majorana materials, Fe(Te,Se)[17] and CaKFe$_4$As$_4$[22].

We first discuss the typical behavior of vortex bound states of *free*Vs in which no ZBCP are observed. We show detailed line-cut measurements across a *free*V in Fig. 2. Similar to what has been observed previously[27], the most pronounced spectral feature is a pair of dispersive vortex bound states (Fig. 2c) at non-zero energies (Figs. 2e-2g). Owing to the high resolution of our data, we further distinguish two additional discrete vortex bound states located at ±2.6 meV and ±3.7 meV, respectively, which display the quantized behavior[33]. Those discrete levels can be observed directly in the raw data (Fig. 2c) and be recognized more clearly in a curvature plot (Fig. 2d). The extraction of the spatial evolution of those vortex bound states is shown in Fig. 2g. By a more detailed analysis (see in the supplementary information), we find that the dispersive and discrete vortex bound states can be attributed to the outmost $d_{xy}$ band and the inner $d_{yz}$ bulk band, respectively. This implies that the topological bands are likely decoupled from the quasiparticle excitations in a *free*V.

To characterize the ZBCPs, we next focus on an *imp*V (Fig. 3a). Through a line-cut measurement across the *imp*V (Fig. 3c), we find that the ZBCPs (Fig. 1f) remain fixed at zero energy (Fig. S1a), with their intensity gradually decreasing to zero when moving away from the vortex center. This non-splitting behavior can be also observed clearly from a waterfall plot (Fig. 3e) and an overlapping plot (Fig. 3f) of the same *imp*V. This phenomenon is similar to the behavior reported for MZMs in the known FeSC Majorana materials[2]. However, unlike the nearly isotropic MZMs discussed previously[2,22], the intensity line profile of ZBCPs across the *imp*V shows considerable asymmetry (Fig. 3g). This asymmetry is intimately connected to the anisotropy of vortex quasiparticle density induced by the impurity. As shown in Figs. 3a and 3b, the intensity maximum of vortex bound state deviates from the center of impurity. This polarization can be distinguished more precisely by comparing the position of the strongest ZBCP in the vortex line-cut and the position of the weakest superconducting coherence peak in the zero field line-cut (Fig. S1e). We marked those two positions as the horizontal dashed bars in Figs. 2c and 2d, respectively. The impurities may disturb the density of Cooper pairs asymmetrically near the vortex core, and lead to an anisotropic vortex. Beside the isolated ZBCPs, a pair of energy-symmetric side peaks display the dispersive behavior across the vortex core. By fitting the *dI/dV* spectrum measured at the vortex center, we find that the full width at half maximum (FWHM) of ZBCP is about 0.5 meV and that of the side peaks is larger than 1 meV (Figs. S1a-S1b). Considering this increased broadening of non-zero energy vortex bound states, the coexistence of well-isolated ZBCP and dispersive high-energy bound states indicates a larger quasiparticle minigap between ZBCP and lowest excitations, but much smaller gaps in between different higher levels of vortex bound states. Thus, the large energy broadening merges the high energy levels to form an apparent dispersive bound state but leaves the zero mode still isolated. It resembles the spectra of topological quantized vortices near the zero-doping-limit[17,34,35] (the case where the Dirac point is at the Fermi level), where the MZM emerges as the zeroth level in the bound states sequence.

To further substantiate our interpretation of the nature of the higher energy dispersive vortex bound state and explain the asymmetry of the zero-energy state, we perform theoretical calculations using a lattice model (see details in supplementary information). The model is based on a 2D proximitized Dirac fermion with a superconducting vortex and an impurity placed off the vortex center. The parameters used in the calculations are based on fitting to the analytical model of Majorana wavefunction[2,8,22,36] (Fig. 3g). With broadening modeled using an effective temperature resulting in FWHM comparable to the measured value (Figs. S1a-S1b), a good agreement with the experimental data is obtained (Fig. 3h). As the chemical potential $\mu \approx 1.9\Delta$, the first non-zero energy state lies at $E_1 \approx 0.5\Delta$ and the following states are closely spaced in energies. Their spacing is smaller than the peak broadening and thus they appear to be a single dispersive state (a corresponding simulation without broadening applied can be found in Fig. S6). Moreover, the presence of a strong impurity introduces asymmetry in the zero-energy state wavefunction (Fig. 3i) consistent with the measurements, significantly increasing the decay length in that direction (the left side in Fig. 3j), while the opposite side remains largely compatible with the analytical solution (the right side in Fig. 3j). The full consistency between the experiment and theoretical simulations strongly support that the observed ZBCP is an asymmetric MZM from the Dirac states near the zero-doping limit[17,34,35].

We now discuss a possible mechanism for the distinct behavior of different types of vortices and the tunability of Majorana modes in LiFeAs. First, we analyze the *free*V cases. As shown in Figs. 4a-4b, which depict the topological bands of LiFeAs known from a previous ARPES measurement[29], in clean regions $\mu$ is located within the Dirac surface states of the TI phase. The puzzle of absence of MZMs in previous measurements of *free*V can now be resolved by noticing the behavior of Dirac surface states observed in the ARPES work[29]. As the dispersion bends back, it tends to form a Rashba-like dispersion in the upper branch of Dirac surface state (Figs. 4a-4b), which leads to the Fermi level crossing the helical Dirac electrons twice. Accordingly, two MZMs emerge in a core of *free*V (middle panel of Fig. 4c). The pairs of unprotected MZMs can fuse with each other immediately to become fermionic bound states at the gap edge (right panel of Fig. 4c). Consequently, the vortex bound states in *free*Vs have no zero modes with their behavior fully explained by the bulk bands (Fig. 2g). It is worth noting that the TI phase can also support single vortex MZM in LiFeAs, if the hole doping can be introduced in the sample that moves the Fermi level to the lower portions of the Dirac cone where the Fermi level crosses the surface state only once.

For the *imp*V cases, we propose that the presence of an impurity can significantly affect its vicinity in several ways. First, the impurity can provide electron doping, lifting $\mu$ above the TI regime. This conjecture is supported experimentally, as demonstrated in Figs. S2-S4 and the supplementary information. As shown in Figs. 4a and 4b, the energy separation between the two Dirac points is only about 15 meV[29], so that the electron doping can easily push the region surrounding the *imp*V into the upper TDS regime. It has been predicted that within the TDS phase two pairs of helical Majorana modes propagating along the vortex line will emerge (middle panel of Fig. 4d). Such a nodal vortex line would exhibit a constant local density of states[30,31] (right panel of Fig. 4d). Furthermore, a recent theory proposed that the nodal vortex line could be gapped out by a $C_4$ symmetry breaking perturbation[31] (left panel of Fig. 4e). In this way, a single MZM could be stabilized on the sample surface (middle and right panel of Fig. 4e). Remarkably, we observe that the impurities of the *imp*V with ZBCPs have an asymmetric shape which indeed induce asymmetric stress, consequently breaking the $C_4$ symmetry[37,38] (inset in Fig. 4b). This effect makes the nodal vortex line to be full-gapped and transforms the helical Majorana modes into a well localized MZM in *imp*Vs. Finally, we note that the impurity must affect sufficiently large surrounding volume, so that the topological vortex line extends into the bulk. This is necessary to allow for adequate spacing between its two ends and in consequence prevents the hybridization of the two emerging MZM located there. This requirement for strong impurity influence is further corroborated by the observation that MZMs are absent in most *imp*Vs pinned to weak impurities (see details in the supplementary information and Fig. S5). In this case, the characteristics of vortex bound states are either similar to the *free*V (Fig. 4c) or the case of nodal vortex line that emerged from full-symmetric TDS phase (Fig. 4d).

Our work demonstrates the first realization of tunable MZMs in a truly homogeneous superconductor, which may bridge the gap between quantum physics and quantum engineering in FeSC Majorana platforms[25]. Although the tunability observed here is enabled by native impurities, other controllable strategies can be explored soon. Since the atom manipulation by an STM tip is routinely performed[39], it could be used for artificial design of MZM patterns in a vortex lattice by moving impurities to selected vortices. We also propose a new tuning scheme that facilitates a fast, on-demand tunable Majorana device that could be fabricated by combining electrostatic gating effect[40] and piezoelectric stress[41] with LiFeAs superconductor (Fig. 4f). The creation and annihilation of MZM could be then electrically controlled with a fast manipulation rate, realizing one of the necessary conditions for non-Abelian Majorana braiding and fault-tolerant topological quantum computation.

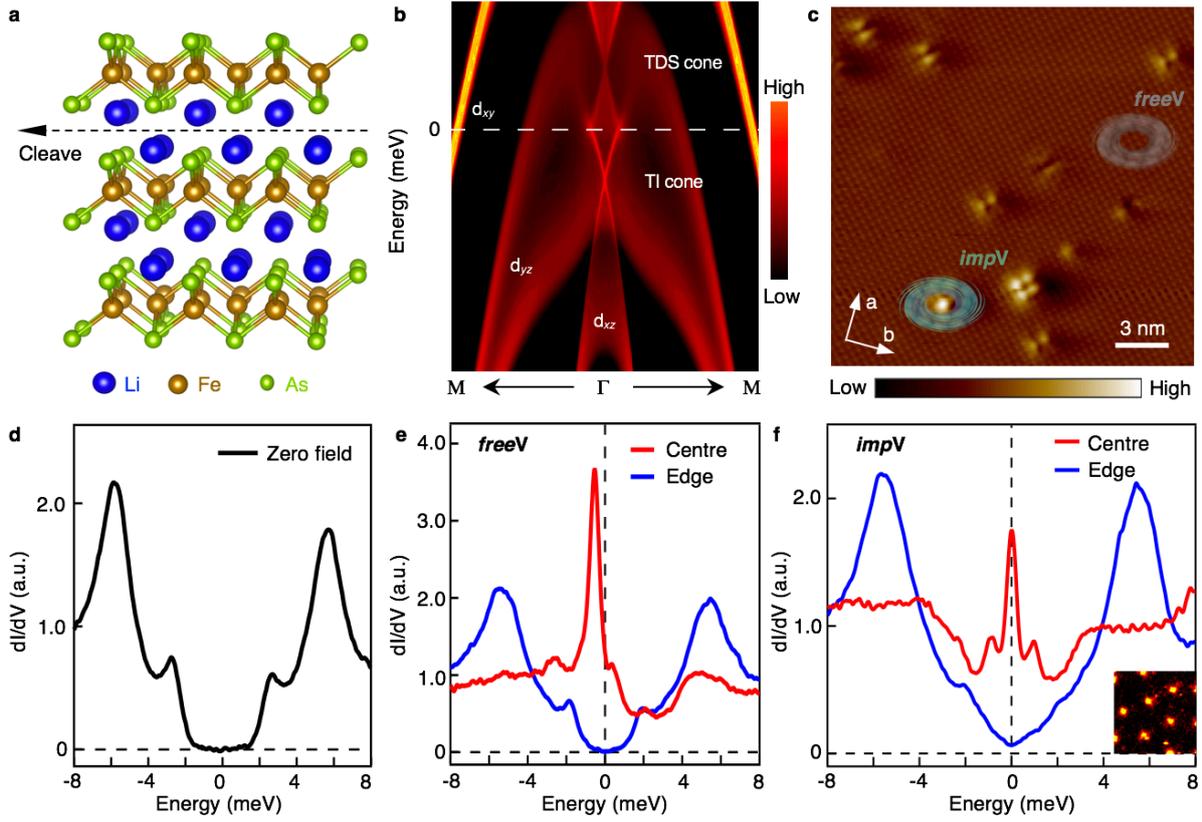

**Fig. 1 | Vortices with and without zero-bias conductance peaks in LiFeAs. a**, Crystal structure of stoichiometric LiFeAs. The natural cleavage plane is indicated by the black dashed line. **b,** (001)-surface-projected band structure of LiFeAs. [Adopted from Fig. 2d of ref. 29]. Two topological cone-like bands appear in the calculation near the Fermi level. The lower one is the Dirac surface state of a topological insulator (TI) phase, and the upper one is the bulk Dirac fermion of a topological Dirac semimetal (TDS) phase. **c,** Atomic resolution STM topography of LiFeAs (scanning area, 20 nm by 20 nm), axes a and b indicate the Fe–Fe bond directions. The native impurities are sparsely distributed on the homogeneous surface which maintains large, well-ordered surface area with its electronic properties unchanged (Fig. S3b-S3c). The vortices on LiFeAs surface belong to two classes: free vortices (*free*V) on the clean surface (gray symbol) and impurity-assisted vortices (*imp*V) pinned to native impurities (green symbol). **d,** A typical tunneling conductance spectrum measured on a clean area of LiFeAs surface under zero field. Two bulk superconducting gaps are identified: $\Delta_1 = 2.7$ meV (due to the outer hole Fermi surface), and $\Delta_2 = 5.8$ meV (due to the inner hole Fermi surface). **e,** Sharp non-zero-energy vortex bound states at the center of a *free*V. **f,** Zero-energy vortex bound states, accompanied by a pair of energy-symmetric non-zero vortex bound states in a typical *imp*V. Inset: the vortex lattice on a zero-bias conductance (ZBC) map under 2.0 T (scanning area, 100 nm by 100 nm). The settings are: sample bias $V_b = -5$ mV; tunneling current $I_t = 200$ pA; and temperature $T_{exp} = 400$ mK.

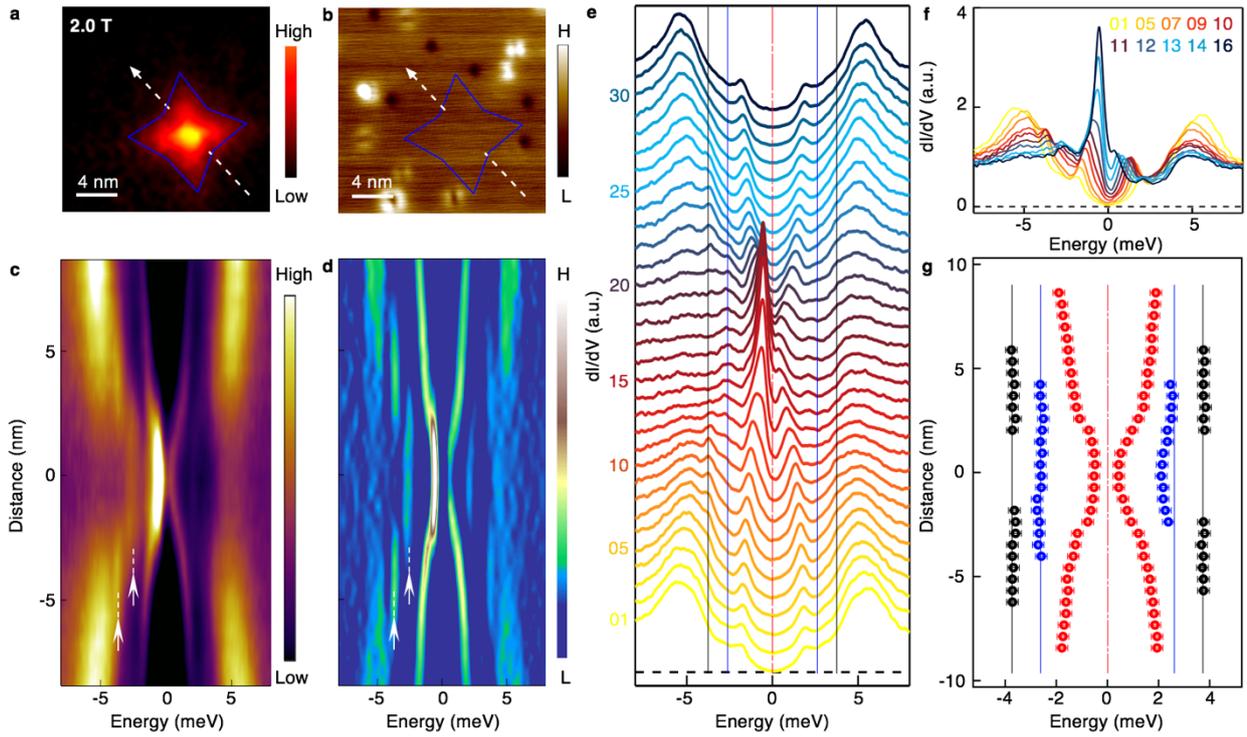

**Fig. 2 | Coexistence of discrete and dispersive vortex bound states in a free vortex. a**, A ZBC map around a *free*V. The zero energy LDOS exhibits star-like shape, with its quasiparticle tails along the nearest As-As directions. **b,** The corresponding topography of **a**. **c,** Line-cut of *dI/dV* measured in the *free*V along the white dashed lines indicated in **a** and **b**. **d,** Curvature intensity plot of **c**. The white dashed lines with arrows indicate the two discrete vortex bound states at high energy. **e,** Waterfall-like plot of **c**. **f,** Plot of several overlapping spectra selected from **e**. **g,** Extracted energy positions of the vortex bound states in **c-e**. Two sets of vortex bound states could be identified, *i.e.* $d_{yz}$ orbital related discrete levels (blue and black symbols), and $d_{xy}$ orbital related dispersive levels (red symbols).

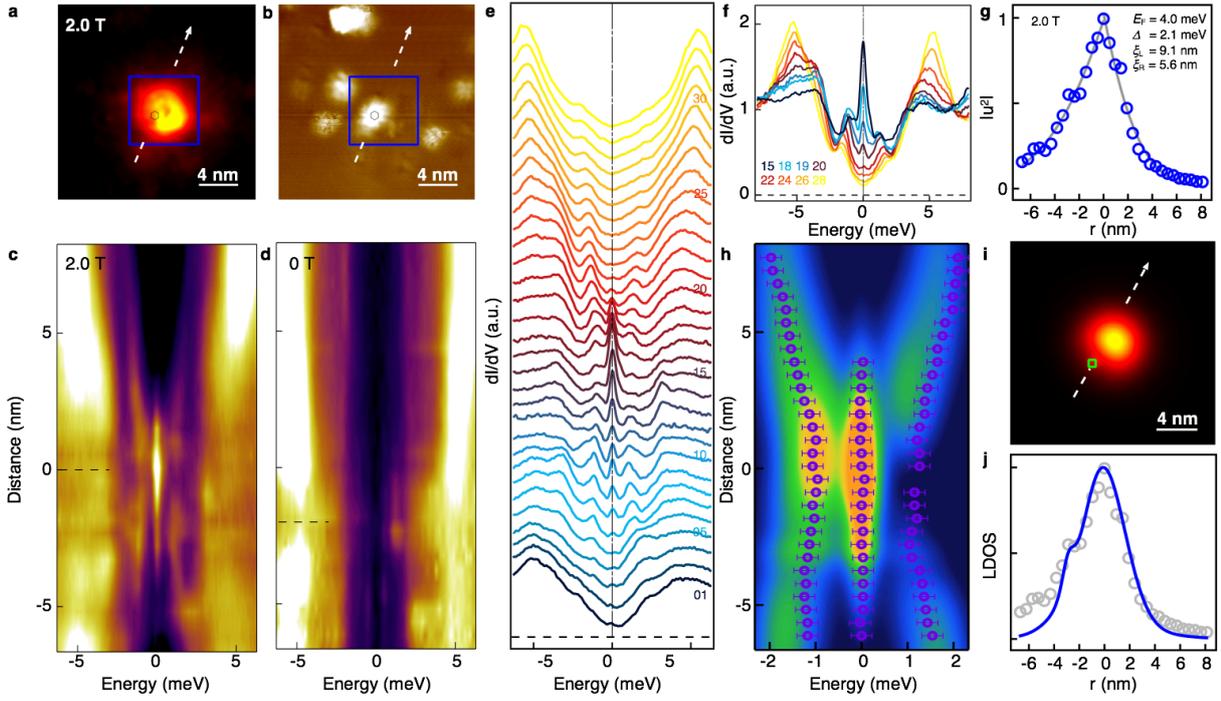

**Fig. 3 | Asymmetric Majorana zero modes in an impurity-assisted vortex. a**, A ZBC map of an *imp*V. **b,** Corresponding STM topography of **a**. Blue boxes mark the same area where the vortex appears, while the black hexagons mark the center of the impurity. **c**, Line-cut of *dI/dV* measured under 2.0 T along the white dashed line indicated in **a**, which demonstrates the spatial evolution of the vortex bound states. **d,** Line-cut of *dI/dV* measured under 0 T and at the same location as in **c**, which demonstrates the spatial evolution of impurity bound states. The black dashed lines in **c** and **d** indicate the center positions of the vortex and the impurity, respectively. **e,** Waterfall-like plot of **c**. **f**, Eight spectra selected from the 2.0 T data in **e**. **g**, Blue symbols: zero-bias *dI/dV* line profile of zero-bias conductance peaks (ZBCPs) measured along the dashed line in **a**. The parameters of underlying topological bands (*i.e.* $\Delta$ = 2.1 meV; $E_F$ = 4.0 meV) are extracted by fitting the data with an analytical Majorana wavefucntion[2,11,22,36] (gray curves). While the parameters of $\Delta$ and $E_F$ used in the fitting of left and right side are same, the fitting parameter of wavefunction decay length of the two sides are different, *i.e.* $\xi_L$ = 9.1 nm; $\xi_R$ = 5.6 nm. **h,** Simulated local density of states line-cut across a vortex with impurity. The purple symbols are the measured energies of vortex bound states shown in **c** and **e**. **i, j,** Simulation of a Majorana wavefunction influenced by an impurity, a two dimensional zero-energy local density of states and its line profile (blue curve) are shown, respectively. The green square in **i** indicates the impurity center. **j** is traced along the white line indicated in **i**. The gray symbols in **j** are the experimental results shown in **g**. The parameters used in simulation of **h-j** are based on the fitting results from **g**: $\Delta$ = 2.1 meV; $E_F$ = 4.0 meV, $\xi_0$ = 3.6 nm. The energy broadening used in the simulation of **h** are 0.5 meV for ZBCPs and 1 meV for other vortex bound states, as measured in this system (Figs. S1a-S1b).

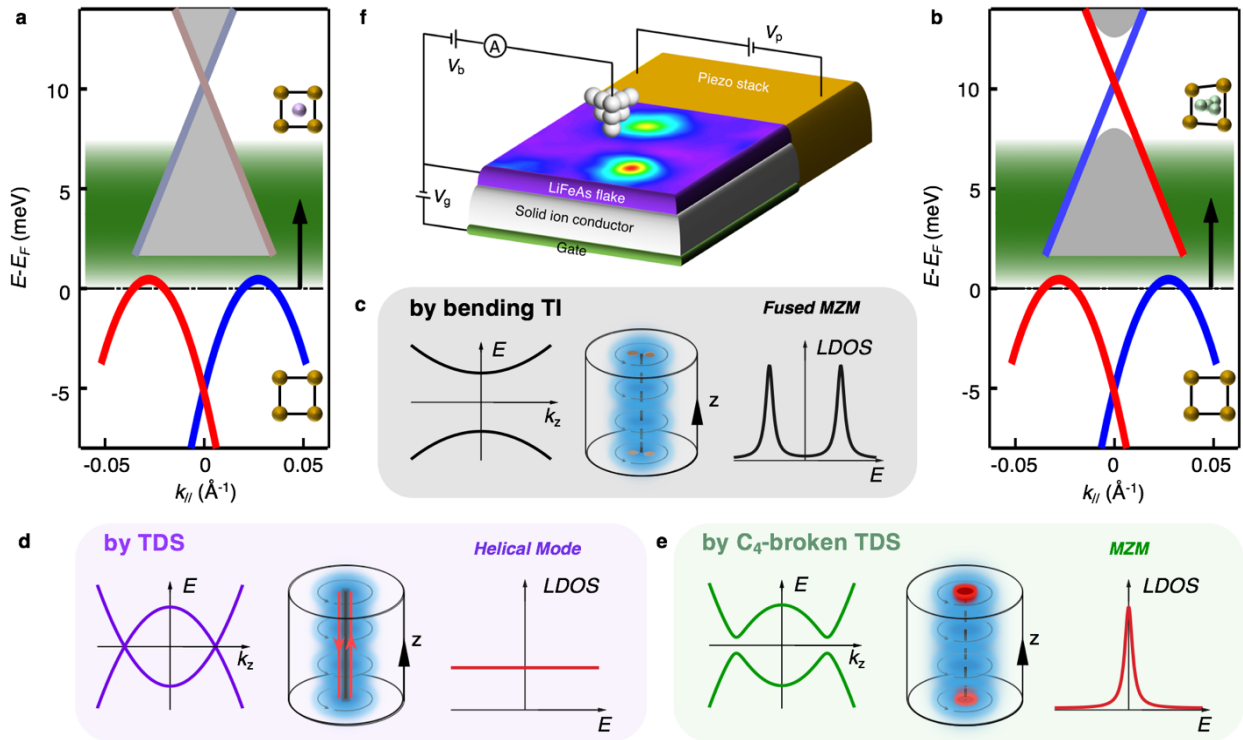

**Fig. 4 | Stabilizing Majorana zero modes by tuning bulk Dirac fermion. a, b,** Schematic depiction of topological band structure of LiFeAs tuned by impurities. Band dispersion of the lower Dirac surface state and the upper bulk Dirac fermion is extracted from ARPES measurements[29]. The Fermi level in the diagram follows from the measured chemical potential ($\mu$) of LiFeAs. The energy separation between the two Dirac points is about 15 meV, the Dirac point of topological surface states is located at about -5 meV. The green shaded region in **a** and **b** indicates the available $\mu$ range tunable by impurities as demonstrated in Figs. S2-S4. Stronger impurity not only tends to induce a larger electron doping effect (Fig. S5a), but also induces a larger asymmetric stress which causes lattice distortion of the bulk and breaks the $C_4$ symmetry of the square lattice near the impurity (inset of Fig. 4b). Thus, the upper bulk Dirac cone remains intact in the region tuned by weaker impurities (**a**), but becomes gapped in the region near the stronger impurities (**b**). **c-e,** The dispersion of the lowest vortex bounds states in the bulk along $k_z$ direction (left panel), the diagram of the vortex line (middle panel), and the local density of states (LDOS) of Majorana modes (right panel) in the cases of the bending TI surface states (near the Fermi level in **a** and **b**), TDS bands which maintains full symmetries (green shaded region in **a**) and $C_4$-symmetry broken TDS bands (green shaded region in **b**), respectively. Majorana modes appear as: fused fermionic bound state, two pairs of mobile helical Majorana modes in the nodal vortex line, and single Majorana zero modes, respectively. **f,** An experimental design to realize a fast, tunable Majorana device using LiFeAs vortex. By combining efficient gating effect (*e.g.* field-effect transistor using solid ion conductors as gate dielectric[40]) and uniaxial deformation (*e.g.* a device controlled by piezo stack which is tightly connected to the sample[41]), the phase transition among trivial vortex bound states, helical Majorana modes and localized MZM could be detected continuously by an *in-situ* STM measurements in the future.

# Supplementary Information for
# Tunable vortex Majorana zero modes in LiFeAs superconductor

Lingyuan Kong[†], Lu Cao[†], Shiyu Zhu[†], Michał Papaj[†], Guangyang Dai, Geng Li, Peng Fan, Wenyao Liu, Fazhi Yang, Xiancheng Wang, Shixuan Du, Changqing Jin, Liang Fu, Hong-Jun Gao[*], and Hong Ding[*]

[†]These authors contributed equally to this work
[*]Correspondence to: dingh@iphy.ac.cn, hjgao@iphy.ac.cn

**Single-crystal growth**

High-quality single crystals of LiFeAs were grown using the self-flux method[26]. The precursor of $Li_3As$ was first synthesized by sintering Li foil and an As lump at about 650°C for 10 h in a Ti tube filled with argon (Ar) atmosphere. Then the $Li_3As$, Fe and As powders were mixed according to the elemental ratio of $LiFe_{0.3}As$. The mixture was put into an alumina oxide tube and subsequently sealed in a Nb tube and placed in an evacuated quartz tube. The sample was heated to 1100°C for 20 h and then slowly cooled down to 750°C at a rate of 2°C per hour. Crystals with a size of up to 5 mm were obtained. To protect the samples from reacting with air or water, all the synthesis processes were carried out in a high-purity Ar atmosphere.

**Scanning tunneling microscopy measurements**

STM/S measurements were conducted in an ultrahigh vacuum (1 × 10$^{-11}$ mbar) USM-1300-$^3$He system with a vector magnet. The energy resolution is better than 0.28 meV. Tungsten tips were calibrated on a clean Au(111) surface before use. To protect the samples from reacting with air or water, LiFeAs sample for STM measurements was mounted in a glove box filled with high-purity Ar atmosphere (> 99.999%). After a quick transfer to STM chamber, it was cleaved *in-situ* at room temperature and transferred to the scanner immediately. Vertical magnetic fields were applied to the sample surface. All data shown in this paper were acquired at 400 mK. STM images were obtained in the constant-current mode. Differential conductance (*dI/dV*) spectra and constant bias maps were acquired by a standard lock-in amplitude at a frequency of 973.0 Hz under a modulation voltage $V_{mod}$ = 0.1 mV. The zero-bias conductance peaks were reproduced in nine impurity-assisted vortices (*imp*Vs) that are measured in four independent samples and with four different tips.

**2D lattice model and theoretical simulation**

To perform the calculations corroborating the analysis of the experimental data we use a tight-binding model of a 2D Dirac fermion described by the following Bogoliubov-de Gennes Hamiltonian:

$$H = \sum_j c_j^\dagger \left( 2v\tau_z s_z - \mu\tau_z + \Delta(j_x, j_y)\tau_x + V_0 \exp\left( -\frac{(j_x - j_{x0})^2 + (j_y - j_{y0})^2}{2\sigma^2} \right) \right) c_j$$
$$+ \left( c_{j+\hat{x}}^\dagger \left( -i\frac{v}{2}\tau_z s_y - \frac{v}{2}\tau_z s_z \right) c_j + c_{j+\hat{y}}^\dagger \left( i\frac{v}{2}\tau_z s_x - \frac{v}{2}\tau_z s_z \right) c_j + H.c. \right)$$

where $j = (j_x, j_y)$ is the index labelling sites on the 2D square lattice, $c_j$ are the annihilation operators acting in Nambu space, $\tau_i, s_i$ are the Pauli matrices acting in particle-hole and spin spaces, respectively, $v$ is the velocity of the Dirac fermion, $\mu$ is the chemical potential,

$$\Delta(j_x, j_y) = \Delta_0 \tanh \frac{\sqrt{j_x^2 + j_y^2}}{\xi_0} \exp(i \arctan \frac{j_y}{j_x})$$

is the superconducting order parameter with a vortex at $j = (0,0)$, $\xi_0$ is the superconducting coherence length, $V_0$ is the strength of an impurity placed at $j_0 = (j_{x0}, j_{y0})$ described by a Gaussian potential. In such a model we calculate the eigenvalues $E_n$ with corresponding wavefunctions that are composed of the particle components $u_{n,\sigma}$ and hole components $v_{n,\sigma}$. With these eigenvalues and wavefunctions we compute the local density of states given by:

$$\rho(E) = -\sum_{n,\sigma} |u_{n,\sigma}|^2 f'(E_n - E) + |v_{n,\sigma}|^2 f'(E_n + E)$$

where $f'(E)$ is the derivative of the Fermi-Dirac distribution. The parameters used in the calculation are: $v = 1, \mu = 0.0475, \Delta_0 = 0.025, V_0 = -0.21, j_{x0} = j_{y0} = -22$. Based on the fit to the analytical wavefunction[11,36] (Fig. 3g), same as the fit performed in refs. 2 and 22, we can translate these values to $\Delta_0 = 2.1$ meV, $\mu = -4$ meV, $\xi = 3.6$ nm. The negative chemical potential ($\mu = -4$ meV) indicates that the underlying topological bands of MZM have hole-like dispersion, which is consistent with our scenario (Figs. 4a-4b). By the eigenvalues and wavefunctions of vortex bound states obtained above, we plot the local density of states along a line-cut through the impurity position (Figs. 3h-3j). To incorporate the extra broadening of vortex bound states observed in Fig. S1a-S1b, we model the broadening using effective temperature of the results with FWHM of 0.5 meV for zero energy mode and 1 meV for the higher energy states, comparable to the measured value.

**Issue of inhomogeneity in FeSC Majorana platforms**

LiFeAs is the most homogeneous material among all the existing topological iron-based superconductors in which vortex MZMs are realized. As for Fe(Te,Se)[2] and (Li,Fe)OHFeSe[16], the topological band structure requires elemental substitutions which induces bulk inhomogeneity; and for another compound CaKFe$_4$As$_4$[22], although it has a stoichiometric bulk, which realizes the fully-theoretical-reproduced spatial patterns of integer-quantized vortex bound states, the cleavage surface is polar and suffers from surface inhomogeneity. Those features complicate the application of MZMs. LiFeAs has a layered structure (Fig. 1a), and it can be cleaved in between the lithium layers, resulting with a charge neutral surface of lithium atoms. Thus, LiFeAs offers a great promise for creating, manipulating, and tuning MZMs, owing to the homogeneous conditions observed both in the bulk and on the surface.

**Origins of vortex bound states in *free*Vs**

To reveal the origin of the two classes of differently behaving vortex bound states in *free*Vs, we measure the vortex shape by zero-bias conductance map. We find that the low energy quasiparticles have a star-like shape with the tails along the Γ-X direction (Fig. 2a). This anisotropy is likely caused by the rounded-square Fermi surface of outmost $d_{xy}$ orbital, where the parallel flat segments of Fermi surface are perpendicular to Γ-X direction[27, 42-47]. Furthermore, it has been resolved clearly in LiFeAs that the larger (smaller) superconducting gap $\Delta_1$ ($\Delta_2$) opens on the inner $d_{yz}$ (outer $d_{xy}$) Fermi pocket, which has smaller (larger) $E_F$[47]. Hence the $d_{yz}$ ($d_{xy}$) orbital related vortex bound states have a larger (smaller) level spacing due to which it is easier (harder) to approach the quantum limit[33], thus appearing as discrete (dispersive) bound states.

**Reproducibility of MZMs observation in *imp*Vs**

We have repeated the observation of spatially non-splitting ZBCPs in 9 different *imp*Vs (two of them are shown in Fig. 3 and Fig. S2) and checked all the necessary aspects carefully. We first determined that the sub-gap states shown in Fig. 3c are not impurity bound states. We measured *dI/dV* spectra under zero field at the same measurement positions as that for the vortex bound states under 2.0 T. This demonstrates clearly that the impurities do not introduce the zero energy bound states (Fig. 3d).

A zero-bias conductance map further shows that the vortex area does not have the zero energy quasiparticles when under zero field (Fig. S1d). Furthermore, we checked the magnetic field evolution of the impurity bound states at the *imp*V location by carefully avoiding vortex pinning. We found the impurity bound states never turn out to be zero energy (not shown). In addition, the observed ZBCPs in an *imp*V are robust against changing tip-sample distance. The ZBCP are stable at the zero energy over two orders of magnitude of tunneling barrier conductance (Fig. S1c), fully consistent with the appearance of a single MZM.

**Evidence of electron doping effect around impurities**

Unlike the well-defined Fermi level in the clean regions[29], the areas with *imp*Vs may have different chemical potential ($\mu$). By measuring the shift of the $d_{xy}$ band top, we reveal that the impurities shift $\mu$ up towards the TDS crossing. It is known that the band top of the outmost $d_{xy}$ orbital appears as a hump in *dI/dV* spectrum at around +33.4 meV (the black curve in Fig. S2a, and Fig. S3). This band top position provides an indicator for $\mu$ variations at different positions. While the band top hump position is fixed across the clean region (Fig. S3), it shifts to lower energies in the vicinity of an impurity (red curve in Fig. S2a), which indicates electron doping induced by the impurity in its surroundings.

We perform detailed measurements of the spatial variation of the band top hump across an *imp*V. A ZBC map for the vortex, topography and line-cut measurement across the vortex are shown in Fig. S2e. To enhance the visualization of the hump and the variations of $\mu$, we perform linear background subtraction on each *dI/dV* spectrum (Fig. S2b). Its corresponding raw data is shown in Fig. S2f. We note that the superconducting gap and even the MZM can be identified in these wide-energy-range measurements. To extract the relative $\mu$ shift reliably, we calculated the negative of the second derivative of the raw data shown in Fig. S2f. Subsequently, the hump energy at each spatial position was defined as the peak energy of negative second derivative spectra, which are extracted by a simple Gaussian fit (Fig. S2g). As shown in Fig. S2b, those extracted hump energies matched with the linear background subtracted intensity very well. We also measure the same line-cut under zero field (Fig. S2c), with the extracted hump energy showing good agreement (Fig. S2d) with 2.0 T results. We have performed the same procedure for all 9 *imp*Vs in this study. Despite the different extent of tunability, we found that all of the impurities shift $\mu$ upward (Fig. S4a). The available $\mu$ range in *imp*V cases is marked as the green shaded region in Fig. 4a and Fig. 4b.

**Evidence of impurity potential dependent tunability**

The scattering strength of impurities can be estimated from the energy of the impurity bound states, *i.e.* a stronger impurity can induce a lower energy impurity bound state[48]. We now turn to study the correlation between the energy of the lowest impurity bound states ($E_{IBS}$) and the multiple tuning effects of impurities. By measuring *dI/dV* line-cut with and without magnetic field at the same positions, we successfully determined several featured quantities of the *imp*Vs simultaneously, such as the $\mu$-shift which is defined as the maximum hump energy deviation from the reference value in a clean region; the energy of the lowest non-zero vortex bound states ($E_{VBS}$) of *imp*Vs ($E_{VBS}$ is also the topological gap which protects MZM); and the $E_{IBS}$ defined above. To reveal the underlying relationships, we measured several *imp*Vs following the strategies mentioned above and summarized them in Figs. S4a and S4b. First, we find that $\mu$-shift is anticorrelated with $E_{IBS}$ in the *imp*Vs (Fig. S4a). It leads to a reasonable conclusion that a stronger impurity induces a greater electron doping in its vicinity. Second, we also found an anticorrelation between $E_{VBS}$ and $E_{IBS}$ (Fig. S2b). It implies that stronger impurities are better for stabilizing MZM in an *imp*V.

The behavior in Fig. S4 points to a possible scenario that stronger impurity introduces larger electron doping, pushing $\mu$ to be closer to the Dirac point of TDS phase. Subsequently, the value of $\Delta^2/E_F$ (for bulk Dirac cone) becomes larger in the vicinity of impurities, which satisfies a prerequisite for larger topological gap of MZM. This impurity-potential-dependent behavior demonstrates the tunability of vortex Majorana modes in LiFeAs.

**The case of non-ZBCP for weak impurities**

In our STM/S experiments, the ZBCP signature of MZMs does not appear in every *imp*Vs. We investigated the statistical distribution of vortices belonging to different classes by measuring all of the vortices in a selected area (150 nm by 150 nm). In this way we have determined that among the *imp*Vs, the probability of observing MZM is 14%. The absence of MZM in other *imp*Vs is likely related to the weakness of impurities to which the vortices are pinned as discussed below.

As summarized in Fig. S4, the degree of tunability due to an impurity is related to its potential strength. A stronger impurity induces a larger chemical potential shift, which leads to a more favorable condition for the emergence of vortex MZMs. Therefore, we note that the lack of MZM in some *imp*Vs is because the given impurity is too weak to sufficiently influence its vicinity. There are several aspects to the effect of the impurity strength. First, for weak impurities there is not enough electron doping in the surrounding region, so the Fermi level is still crossing the bent, Rashba-like part of TI surface state regime, forming two Fermi pockets. In this case, the vortex bound states in the *imp*V behave similarly to a *free*V. Second, a medium strength impurity can elevate $\mu$ to the TDS phase, however there is not enough asymmetric stress induced in its vicinity, which cannot break the $C_4$ symmetry of the square lattice. In this case, two pairs of helical Majorana modes emerge from the fully symmetric TDS phase. The featureless LDOS of helical Majorana modes, emerging from the $p_z/d_{yz}$ bulk band that forms the bulk Dirac fermions, overlaps with a pair of non-zero energy dispersive vortex bound states, which are related to the underlying $d_{xy}$ bulk bands[29]. Thus, the spectrum still appears to be similar to that in a *free*V, but with featureless helical Majorana modes hidden behind it. Finally, in the case of weak impurities the sphere of their influence is limited, so the shift of the chemical potential is restricted to a small region very close to the impurity. This does not allow for a formation of a sufficiently long topological vortex line and the second Majorana mode possibly emerges too close to the surface, fusing with the Majorana located there and in consequence removing the zero-bias conductance peak from the spectrum.

**Example of weak *imp*Vs without MZM**

As shown in Fig. S5, we perform detailed STM/S measurements on a weak *imp*V without MZM. The line-cut measured under the zero field shows that the impurity bound states appear just at the edge of the superconducting gap (Figs. S5d-S5e). It indicates the impurity is quite weak. Across the vortex center, we measure a line-cut intensity plot of the vortex. As shown in Fig. S5c, the spatial evolution of vortex bound states is very similar to that in a *free*V shown in Fig. 2. Selected *dI/dV* spectra are shown in Fig. S5f. We find the coupling of the weak impurity in *imp*V does not change the main features of vortex bound states, although the intensity of vortex bound states is reduced in the *imp*V as compared to the *free*V shown in Fig. 2f, which is plotted with the same scale. Furthermore, we performed the $\mu$-shift measurement on this *imp*V (Figs. S5g-S5h) by the same methods introduced in Fig. S2. We observe that the $\mu$-shift here is smaller (2 ± 1 meV). Those observations are fully consistent with the scenario that the insufficient strength of weak impurities cannot result in the appearance of MZM in weak *imp*Vs.

**Acknowledgements** We thank X.-X. Wu, P. Zhang and J.-Q. Lin for helpful discussion. This work at IOP is supported by grants from the National Natural Science Foundation of China (11888101, 61888102, 51991340, 11820101003, 11921004), the Chinese Academy of Sciences (XDB28000000, XDB07000000) and the Ministry of Science and Technology of China (2016YFA0202300, 2019YFA0308500, 2018YFA0305800 and 2018YFA0305700). The work at MIT is supported by DOE Office of Basic Energy Sciences, Division of Materials Sciences and Engineering under award no. DE-SC0019275.

**Author contributions** H.D. and H.-J.G. designed the experiments. S.Z. and L.C. performed the STM experiments with assistance from L.K., G.L., P.F., W.L., F.Y. and S.D. M.P. and L.F. provided theoretical models and simulations. G.D., X.W. and C.J. provided samples. L.K. and L.C. analyzed experimental data with input from all other authors. L.K. plotted figures with input from all other authors. L.K., H.D. and M.P. wrote the manuscript with input from all other authors. All the authors participated in analyzing experimental data, plotting figures, and writing the manuscript. H.D. and H.-J.G. supervised the project.

**Competing interests** The authors declare no competing interests.

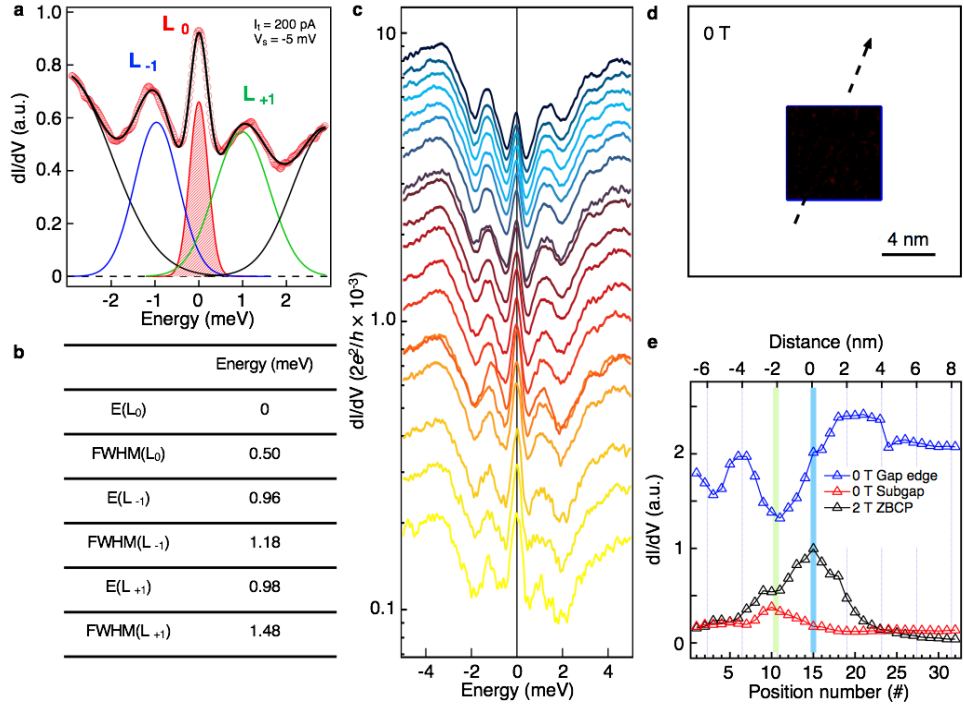

**Fig. S1 | Spectroscopic details of an impurity-assisted vortex. a**, Multi-peak Gaussian fit of the *dI/dV* spectrum measured at the center of the *imp*V shown in Fig. 3. **b,** Fitting parameters from **a**. **c,** Tunneling barrier strength evolution of **a**. The ZBCPs are stable at zero energy for over two orders of magnitude of tunneling barrier conductance. **d,** ZBC map on the area enclosed by the blue box indicated in Figs. 3a and 3b under zero field. It shows that the impurity does not introduce zero-energy impurity bound states in this area. **e,** Line profiles measured along the line indicated in **d** and Figs. 3a and 3b, which demonstrates the mismatch between vortex center (blue bar) and impurity center (green bar). The blue curve is the coherence peak line profile of the integrated intensity of *dI/dV* from -6 meV to -4.5 meV under zero field. The red curve is the subgap line profile of the integrated intensity of *dI/dV* from -1.8 meV to 1.8 meV under the zero field. The black curve is a replot of the zero-bias *dI/dV* intensity line profile under 2.0 T as shown in Fig. 3g.

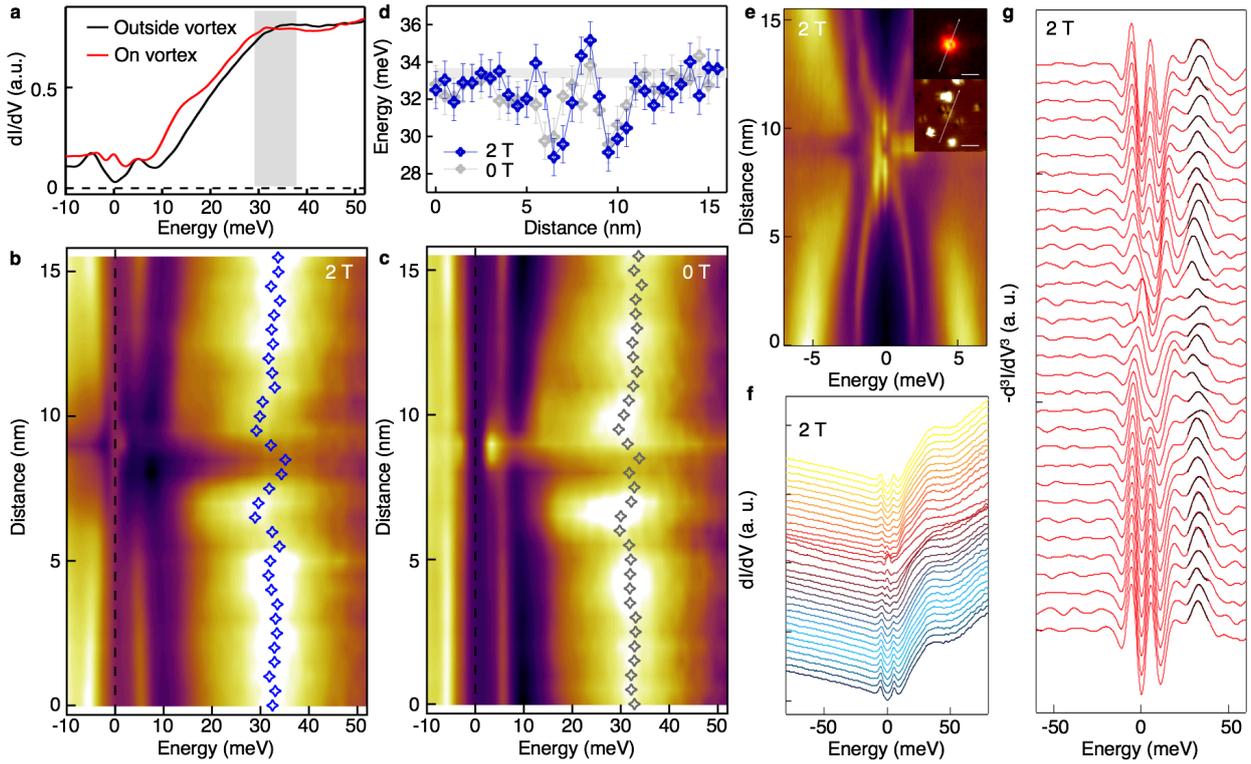

**Fig. S2 | Evidence of local electron doping around an impurity-assisted vortex. a**, Wide range *dI/dV* spectra measured at an *imp*V (red curve) and on a clean surface region without impurities (black curve). The gray shaded region emphasizes the shift of the $d_{xy}$ band top, indicating electron doping around the *imp*V. **b**, Wide range line-cut intensity plot with a linear background subtraction for an *imp*V (see **e**, for basic information). The blue symbols are extracted energies of the band top by the method introduced in **f** and **g**. **c**, Same as **b**, but measured under zero field. **d**, Comparison of spatial evolution of the band top energies across the *imp*V measured under 2.0 T and 0 T. The gray thick line indicates the reference band top energy measured on a clean area without impurities (Fig. S4). The error bars are around 1 meV, determined by the maximum spatial variation of the reference energy ($E_{BT}$ = 33.4 meV) measured in Fig. S4a. **e**, Short range line-cut intensity plot of the *imp*V. Inset: corresponding ZBC map and topography (scale bar: 5 nm). **f** and **g**, Numerical method for the $\mu$-shift extraction. **f**, Raw data of the wide range scan shown in **b**. The $\mu$-shift across the *imp*V is visible in the spectra overlapping plot. **g**, Negative second derivative of the spectra shown in **f**. The signal of the peak in *dI/dV* spectra is enhanced as a peak in the negative second derivative spectra. In order to establish the $\mu$-shift, we extract the energy of the $d_{xy}$ band top by a simple Gaussian fit. The $\mu$-shift in this *imp*V is determined to be 4.5 ± 1 meV.

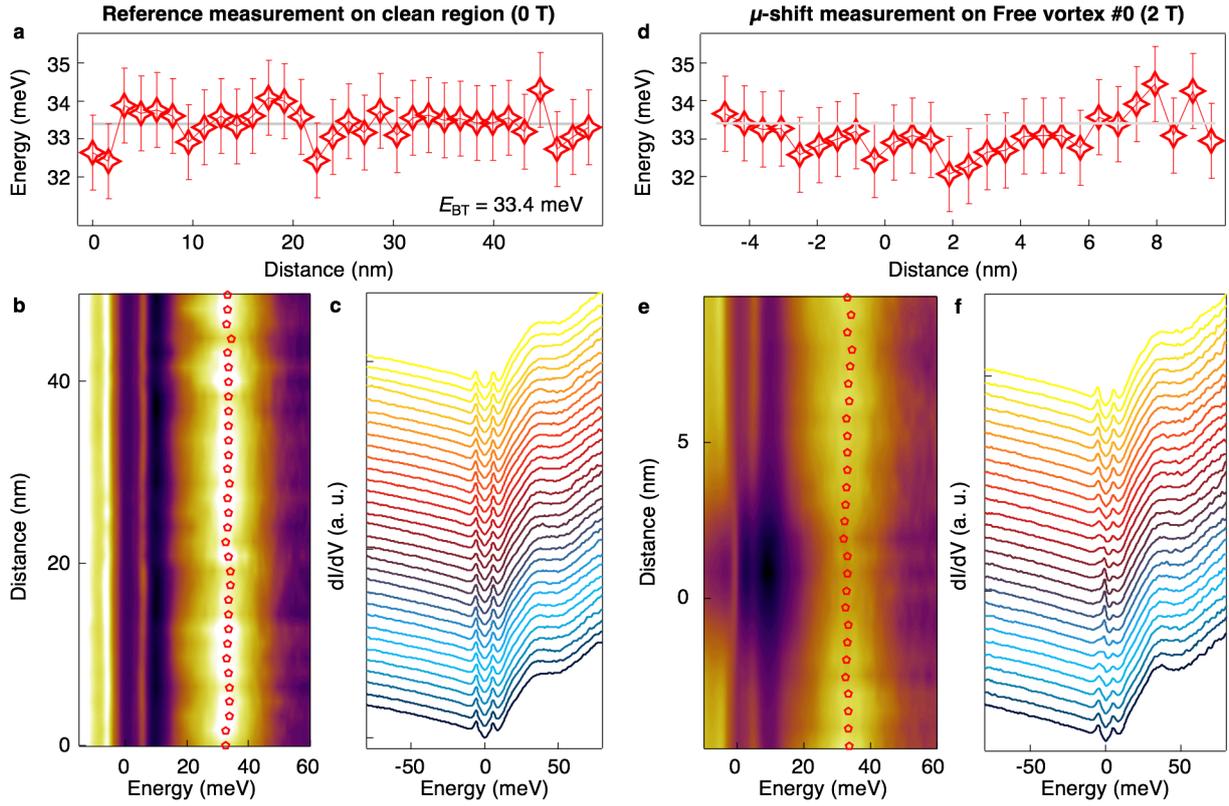

**Fig. S3 | Reference energy of band top in a clean surface region. a,** Spatial evolution of band top energy across a long distance in a clean region without impurities and under zero field. The extraction method used here is introduced in Figs. S2f and S2g. **b,** Intensity plot with background subtracted. The red symbols are the same points as shown in **a**. **c,** Waterfall-like plot of raw data without linear background subtraction. **d-f,** Same as **a-c** respectively, but measured across a *free*V and under 2.0 T. The extracted reference band top energy ($E_{BT}$) is about 33.4 meV, the error bar (1 meV) is determined by the spatial deviation of the singularity energy in a clean surface region.

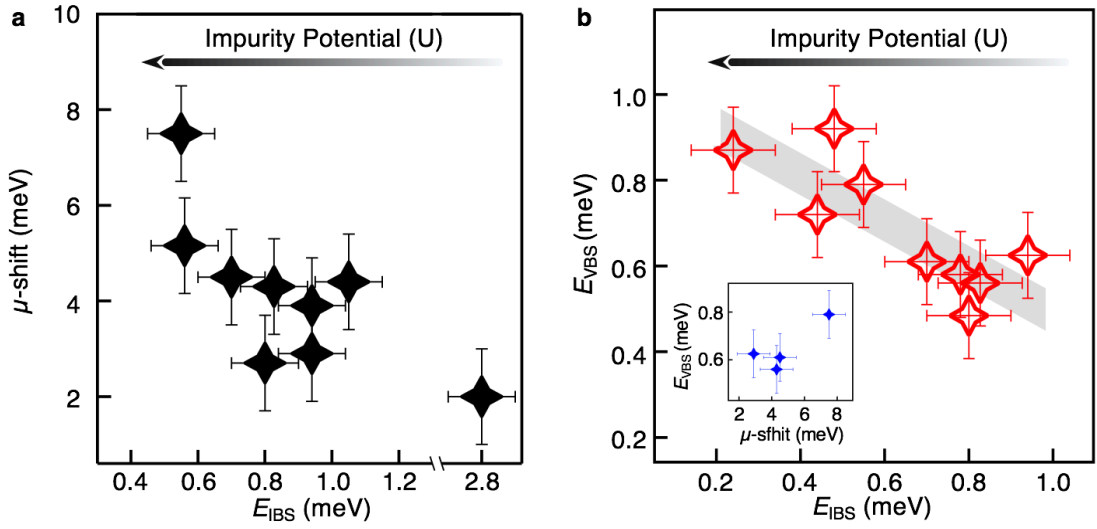

**Fig. S4 | Relationship between vortex tunability and impurity potential. a,** Summary of the relationship between the chemical potential shift ($\mu$-shift) and the energy of the lowest impurity bound states ($E_{IBS}$) among 9 *imp*Vs (some *imp*Vs collected in **a** with weak impurities do not have MZMs). $\mu$-shift is extracted as the maximum band top energy deviation at the *imp*Vs from the reference energy ($E_{BT}$). $E_{IBS}$ is extracted from line-cut intensity measurements under zero field, and along the same line as the measurements under 2.0 T in each vortex. **b,** Summary of the relationship between the energy of the lowest non-zero energy vortex bound states ($E_{VBS}$) and the $E_{IBS}$ among 9 *imp*Vs which have MZMs. $E_{VBS}$ is also the topological gap that determines the protection of MZM. Inset: a summary of the relationship between $E_{VBS}$ and $\mu$-shift among 4 *imp*Vs for which both quantities are measured. A smaller $E_{IBS}$ indicates a stronger impurity potential. The behavior shown in **a** and **b** indicates that vortices pinned to stronger impurities simultaneously have larger $\mu$-shift and larger $E_{VBS}$. It implies that impurity-induced local electron doping shifts $\mu$ towards the TDS point, which is further supported by the data shown in the inset of **b**. The error bars of $E_{VBS}$ and $E_{IBS}$ are 0.1 meV.

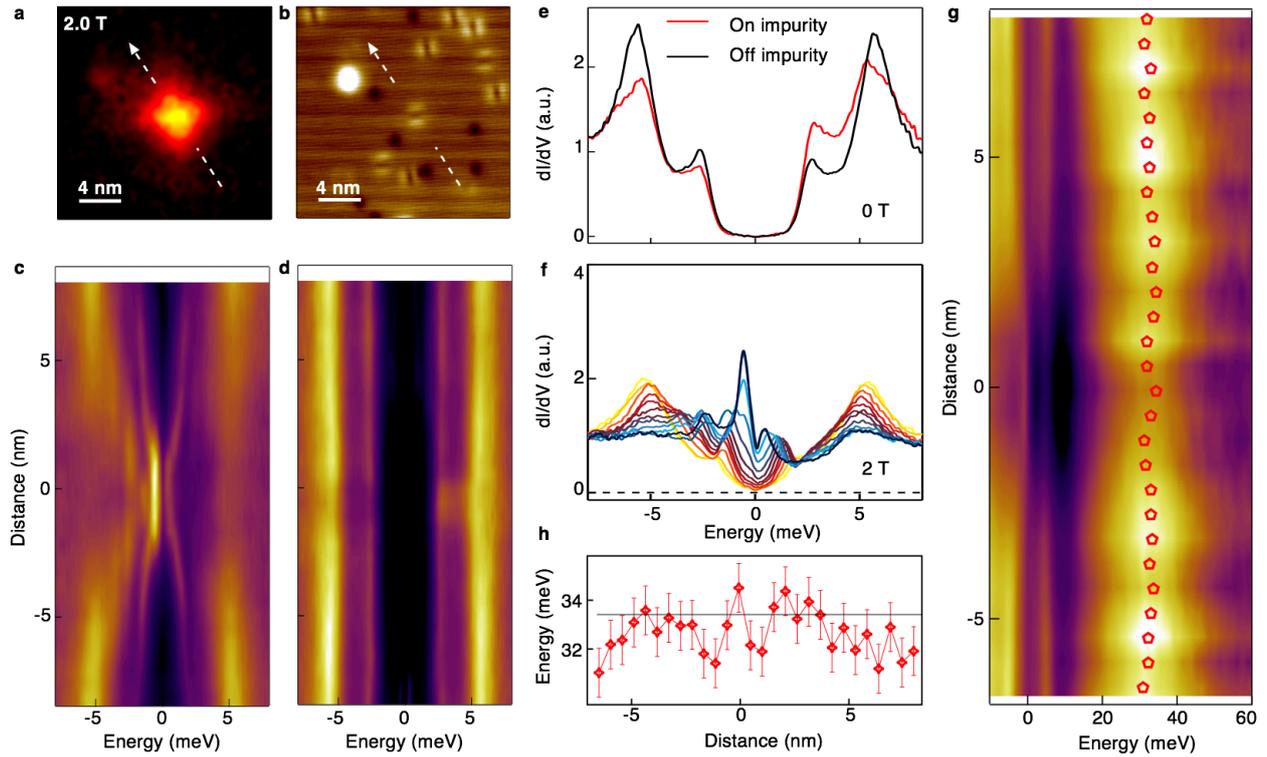

**Fig. S5 | Ordinary vortex bound states in a weak impurity-assisted vortex. a**, ZBC map around an *imp*V. **b,** Corresponding STM topography of **a**. The vortex is pinned to a weak impurity that is formed by D$_{2X}$ symmetric defects located on lattice sites. **c**, Line-cut intensity plot measured under 2.0 T along the white line indicated in **a**, which demonstrates the spatial evolution of vortex bound states. **d,** Line-cut intensity plot measured at 0 T and the same location as that in **c**, which demonstrates the spatial evolution of impurity bound states. **e,** Zero field *dI/dV* spectra measured at and away from the impurity site. **f,** Selected spectra from **c**. **g,** Wide range line-cut intensity plot of the *imp*V with a linear background subtraction. The red symbols are the extracted energies of the band top following the method introduced in Fig. S3. **h,** Spatial evolution of the band top energy across the *imp*V.

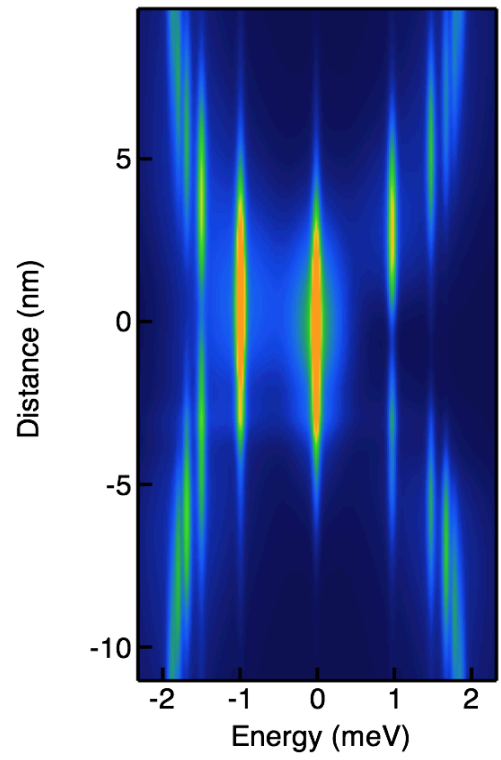

**Fig. S6 | Model calculation of vortex bound states without applying extra broadening.** Without broadening, the non-zero energy bound states can be distinguished and no longer appear as a single dispersive peak in the spectrum.